\documentclass[aps,prl,preprint,groupedaddress]{revtex4-1}

\usepackage{amsmath,amssymb,amsfonts}
\usepackage{graphicx}
\usepackage{hyperref}

\begin{document}

\title{New measurement of $^{12}$C+$^{12}$C fusion reaction at astrophysical energies}
\author{W. P. Tan}
\email{wtan@nd.edu}
\author{A. Boeltzig}
\author{C. Dulal}
\author{R. J. deBoer}
\author{B. Frentz}
\author{S. Henderson}
\author{K. B. Howard}
\author{R. Kelmar}
\author{J. J. Kolata}
\author{J. Long}
\author{K. T. Macon}
\author{S. Moylan}
\author{G. F. Peaslee}
\author{M. Renaud}
\author{C. Seymour}
\author{G. Seymour}
\author{B. Vande Kolk}
\author{M. Wiescher}

\affiliation{Department of Physics and Institute for Structure and Nuclear Astrophysics (ISNAP), University of Notre Dame, Notre Dame, Indiana 46556, USA}

\author{E. F. Aguilera}
\author{P. Amador-Valenzuela}
\author{D. Lizcano}
\author{E. Martinez-Quiroz}

\affiliation{Departamento de Aceleradores, Instituto Nacional de Investigaciones Nucleares, Apartado Postal 18-1027, Codigo Postal 11801, Mexico, D.F., Mexico}

\date{\today}

\begin{abstract}
Carbon and oxygen burning reactions, in particular, $^{12}$C+$^{12}$C fusion, are important for the understanding and interpretation of the late phases of stellar evolution as well as the ignition and nucleosynthesis in cataclysmic binary systems such as type Ia supernovae and x-ray superbursts. A new measurement of this reaction has been performed at the University of Notre Dame using  particle-$\gamma$ coincidence techniques with SAND (a silicon detector array) at the high-intensity 5U Pelletron accelerator. New results for $^{12}$C+$^{12}$C fusion at low energies relevant to nuclear astrophysics are reported. They show strong disagreement with a recent measurement using the indirect Trojan Horse method. The impact on the carbon burning process under astrophysical scenarios will be discussed.
\end{abstract}

\pacs{}
\keywords{${}^{12}$C+${}^{12}$C, fusion reaction, carbon burning, particle-gamma coincidence technique, St. ANA accelerator}

\maketitle

\section{Introduction}

The main nucleosynthesis products of stellar helium burning are the $^{12}$C and $^{16}$O isotopes. For massive enough stars, the subsequent phases of stellar evolution are dictated by fusion reactions such as ${}^{12}$C+${}^{12}$C and ${}^{12}$C+${}^{16}$O \cite{hoyle1954,wiescher2012,pignatari2013}. The cross sections of both fusion reactions are characterized by significant uncertainties associated with the possible emergence of low energy resonances but also a significant suppression due to the so-called hindrance effect which is suggested to reduce the cross section due to the incompressibility of nuclear matter in the collision event \cite{gasques2007}. A dramatic increase in the fusion rate of ${}^{12}$C+${}^{12}$C as suggested recently \cite{tumino2018} may significantly change the abundance distribution in oxygen-neon white dwarfs and in the burning patterns of massive stars in their evolution to core collapse supernovae \cite{pignatari2013}.

Type Ia supernovae (SN) are interpreted as the consequence of explosive carbon burning ignited near the core of the white dwarf star in a binary system \cite{hillebrandt2000}. The ${}^{12}$C+${}^{12}$C fusion process is supposed to be the dominant energy source for pre-ignition processes such as carbon simmering and the ignition itself \cite{martinez-rodriguez2017}. However the ${}^{12}$C+${}^{16}$O reaction may also play a significant role depending on the associated fusion rates \cite{fang2017} and the environmental conditions such as ${}^{16}$O abundance, temperature, and density \cite{gasques2007,martinez-rodriguez2017}. Recent studies showed indeed that the ${}^{12}$C+${}^{16}$O rate is expected to have an unusually large effect on the calcium and sulfur yields in SN Ia, e.g., the higher ${}^{12}$C+${}^{16}$O rate suppresses the alpha-particle abundance, which in turn decreases the Ca/S ratio \cite{martinez-rodriguez2017}.

X-ray superbursts, another phenomenon involving binary compact star systems, are thought to be ignited by the carbon fusion reactions in the burning ashes of accumulated hydrogen and helium on the surface of accreting neutron stars \cite{brown1998, brown2004}. For such an ignition condition of unstable burning, the mass fraction of $^{12}$C has to be at least above 10\% in the ocean of heavy ashes accumulated from previous rp-process burning of X-ray bursts \cite{cumming2001}. However, X-ray burst models could not produce a high enough carbon abundance with known nuclear physics \cite{schatz1999,cyburt2016}. The uncertainty of the rate of the X-ray burst trigger reaction $^{15}$O($\alpha,\gamma$) \cite{tan2007} may reduce the tension a little but certainly not enough \cite{fisker2007}. To make superburst models work, a hypothetical resonance at 1.5 MeV of the center of mass fusion energy of $^{12}$C+$^{12}$C was suggested \cite{cooper2009}.


In the following sections, we will discuss first the status of the $^{12}$C+$^{12}$C fusion cross section data followed by a presentation of the new experimental data obtained at the University of Notre Dame in comparison with previous results.
 
\section{Current Status of $^{12}$C+$^{12}$C}

Extensive efforts, both experimentally and theoretically, have been invested in the determination of the $^{12}$C+$^{12}$C reaction rate for all associated reaction channels. Despite these efforts, large uncertainties remain in the reaction rate especially when extrapolating the data into the astrophysically important energy range (the Gamow window) \cite{pignatari2013}. The predicted rates depend sensitively on adopted model parameters, hindrance effects, and the possibility of cluster, dynamic or molecular resonances at relevant energies \cite{cindro1988, gasques2007, jiang2007, diaz-torres2018}.

Extending and improving the quality of experimental data towards lower energies is therefore crucial for reducing the uncertainties, giving more robust extrapolation towards lower energies, and ultimately providing more reliable reaction rates for the study of carbon burning in stars and other stellar environments.

Astrophysically, the most important energy range for the carbon fusion cross section is about $1-3$ MeV in the center of mass. This is a very challenging range for direct measurements due to the dramatic reduction of the cross section by the Coulomb barrier. The three main channels of $^{12}$C($^{12}$C,p)$^{23}$Na ($Q=2.241$ MeV), $^{12}$C($^{12}$C,$\alpha$)$^{20}$Ne ($Q=4.617$ MeV), and $^{12}$C($^{12}$C,n)$^{23}$Mg ($Q=-2.598$ MeV) reactions can populate the ground state or excited states in the respective residual nuclei that subsequently decay by gamma emission to the ground state.

Earlier direct measurement of the n-emission channel near astrophysically relevant energies conducted at Notre Dame demonstrated that this channel contributes less than 5\% to the total reaction rate \cite{bucher2015}, similar to the case of the $^{27}$Si+n channel in the $^{12}$C+$^{16}$O fusion reaction \cite{fang2017}.

\begin{figure}[htb]
\centering
\includegraphics[width=0.5\textwidth]{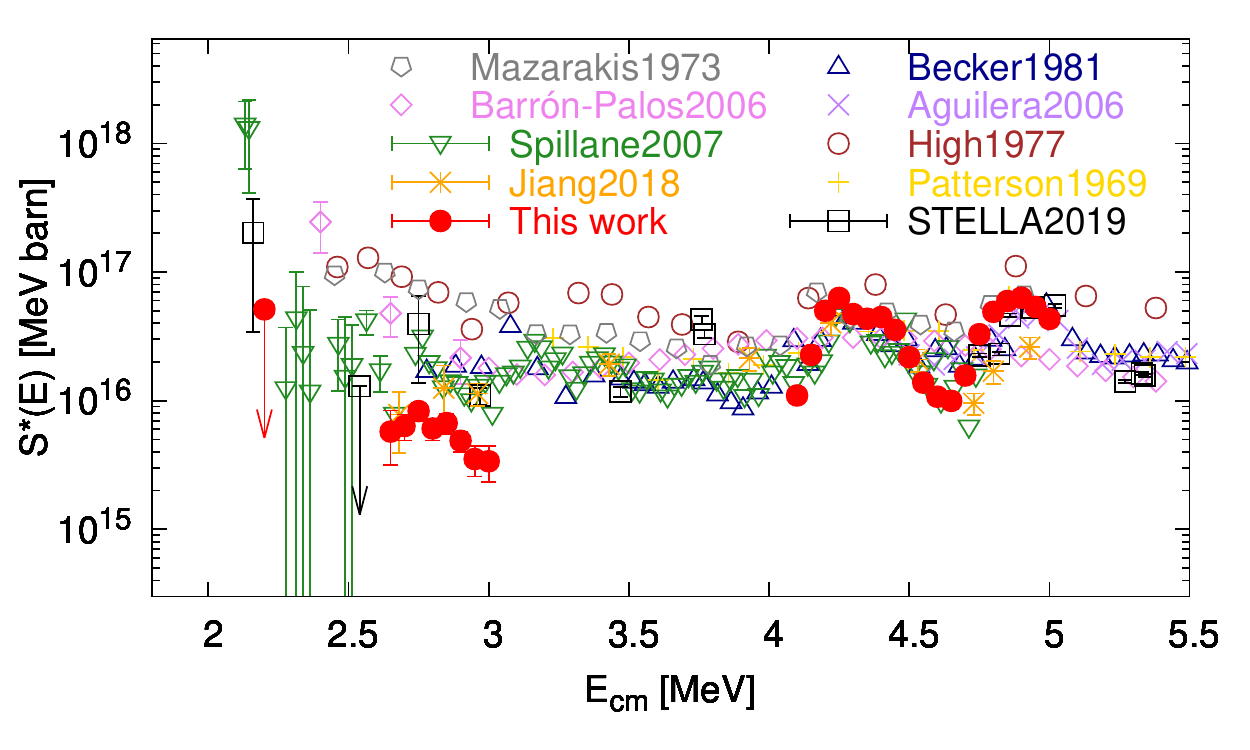}
\caption{
\label{fig_totsfac}
The total $S^*(E)$ factor for this work is shown and compared with previous direct measurement data \cite{patterson1969,mazarakis1973,high1977,becker1981,aguilera2006,barron-palos2006,spillane2007,jiang2018,stella2019}. See the electronic version for better presentation of the details.
}
\end{figure}

Most of the early experimental efforts following the observation of resonances in carbon fusion cross sections by the Chalk River experiment \cite{almqvist1960} are direct singles measurements with detection of either charged particles \cite{patterson1969,becker1981,mazarakis1973,zickefoose2018} or gamma radiation \cite{spillane2007,aguilera2006,satkowiak1982,high1977,kettner1980,dasmahapatra1982,barron-palos2006}, which are shown in Fig. \ref{fig_totsfac}. Such an approach may suffer background issues at low energies due to the rapidly declining cross section. For this reason coincidence techniques between the particle and subsequent $\gamma$ transitions have been applied for better identification of the specific decay patterns of the $^{24}$Mg compound system. The most important channels for coincidence measurements are the $p_1$ transition (emission of protons to the first excited state in $^{23}$Na followed by the 440 keV $\gamma$ transition to the ground state) as well as the $\alpha_1$ transition (emission of alpha particles to the first excited state in $^{20}$Ne with its subsequent ground state decay via the 1634 keV $\gamma$ transition).


An early test experiment using the particle-$\gamma$ coincidence technique for the ${}^{12}$C+${}^{12}$C reaction was conducted at Argonne National Laboratory \cite{jiang2012}. Similar techniques were then used to measure the cross sections at a few points with energies below $E_{\text{cm}}\le 5$ MeV (labeled as ``Jiang2018'' \cite{jiang2018} and ``STELLA2019'' \cite{heine2018,stella2019} in Fig. \ref{fig_totsfac}). The limited number of data points of these experiments suggests a smooth excitation function but can not exclude the resonance structures suggested by other works. The supposedly ``thin'' targets used in the previous works (e.g., \cite{becker1981,jiang2018}) were not thin enough resulting in the extracted reaction yield averaged over the large energy loss range of a few hundred keV in the target due to the large stopping power of the carbon beam \cite{fowler1948}, which will be discussed later. In addition, there is some uncertainty in the target thickness, owing to carbon buildup on the targets during beam bombardment that can easily cause significant changes of the thickness.

A more recent measurement using the indirect Trojan Horse method (THM) provided an additional set of data for the low energy range of the $^{12}$C+$^{12}$C fusion process \cite{tumino2018}. It extracted a number of resonances between $E_{\text{cm}}=0.8-2.7$ MeV predicting a significantly higher reaction rate than all previous estimates. However, there is some concern about the analysis of the THM data, in particular with respect to the treatment of Coulomb corrections \cite{mukhamedzhanov2019}. Coulomb penetrability calculations suggest that the extracted THM resonance strengths are larger than allowed for carbon cluster configurations. Not included in the conversion of the THM data to reaction cross section is the possibility of the hindrance effect \cite{jiang2007}. While the existence of the hindrance effect is not yet finally confirmed, it increases the uncertainty of the final reaction rate. THM measurements do not result in absolute cross sections but are typically normalized to existing experimental data. Therefore, the extracted cross sections towards lower energies depend very sensitively on the theoretical analysis as well as on the normalization process. 

In the following, we will present a direct measurement of the $^{12}$C+$^{12}$C reaction cross sections based on the particle-$\gamma$ coincidence technique and a target approach that allows us to determine the yield from thin target layers through a differential thick target analysis that can potentially address some of the uncertainties present in previous experiments.

\section{Experimental Setup}

In this work both charged particles (i.e., protons and alphas) and $\gamma$-rays emitted from the $^{12}$C+$^{12}$C fusion process were measured simultaneously. A beam of $^{12}$C$^{2+}$ and $^{12}$C$^{4+}$ ions (up to about 13 particle $\mu$A) with $E_{cm}=2.2-5$ MeV (where uncovered energy gaps are due to limited beam time) was produced by the single-ended 5U Pelletron accelerator at the Nuclear Science Laboratory (NSL) of the University of Notre Dame. This accelerator provides high intensity (tens of particle $\mu$A) heavy ion beams, up to ${}^{40}$Ar. The target was a highly ordered pyrolytic graphite (HOPG)
 \cite{soldano2010}, which has a layered structure of multiple thin graphene sheets \cite{novoselov2005}. The advantage of using HOPG as target material is its superior purity compared to natural graphite. Impurity of hydrogen and deuterium in the target can
cause background in charged particle spectra \cite{morales-gallegos2018,zickefoose2011a} while impurity of $^{23}$Na can severely affect the gamma spectra as discussed later.
The HOPG target with a dimension of $2\,$cm$\times2\,$cm$\times1\,$mm served also as a water-cooled beam stop. For accurate reading of the beam current, permanent magnets and a negative suppression voltage of 1500 volts was used around the target.

The Silicon-detector Array at Notre Dame (SAND) \cite{fang2017} for detection of protons and alpha particles consists of six YY1-type silicon detectors and one S2-type silicon detector \cite{micron}, covering polar angles from 102${}^{\circ}$ to 146${}^{\circ}$ and 151${}^{\circ}$ to 170${}^{\circ}$ in the laboratory frame. Each wedge-shaped YY1 is segmented into 16 strips on the front junction side with six YY1 detectors forming a ``lampshade'' configuration. The CD-shaped S2 detector is double-sided and has 48 rings on the front junction side and 16 segments on the back ohmic side. The solid angle covered by the detectors is 4.1\% of 4$\pi$ for each YY1, and 5.4\% of 4$\pi$ for the S2, as determined from measured and design dimensions in addition to an alpha source calibration. For the measurement of the $\gamma$ rays, a HPGe detector with relative efficiency of 109\% (relative to that of $3"\times 3"$ NaI at 1.33 MeV) was placed in a 10-cm thick lead castle and positioned right behind the target to maximize the detection efficiency of $\gamma$ rays. Radioactive sources of $^{7}$Be, $^{56}$Co, $^{60}$Co, $^{66}$Ga, $^{133}$Ba, $^{137}$Cs, and $^{152}$Eu were used to calibration the HPGe detector for an energy range of $0.1-4.8$ MeV. And the absolute $\gamma$ peak efficiency was determined to be 2.30\% at 440 keV and 1.22\% at 1634 keV with an uncertainty of about 5\%. The data were collected by the VMUSB data acquisition system implemented at NSL, where 160 channels of signals from the silicon detector array were processed via an ASIC (Application Specific Integrated Circuit) readout system. The core component of the system is HINP16C, a 16-channel ASIC specifically developed for readout of silicon strip detectors used in low- and intermediate-energy heavy-ion reaction experiments \cite{engel2007}. The HPGe detector was read out by a 13-bit high resolution ADC (MADC-32) from Mesytec \cite{mesytec}. More details of the setup can be seen in the earlier study on the ${}^{12}$C+${}^{16}$O reaction \cite{fang2017}.

\section{Data analysis and results}

Cross sections for various exit channels in the $^{12}$C+$^{12}$C fusion reaction were obtained from the thick-target yields for $\gamma$ transitions associated with the different decay channels and protons/alpha-particles in coincidence. The same analysis procedure for all beam energies is presented as follows.

The thick-target reaction yield is obtained from the number of events detected per incident carbon nucleus on the target for a given reaction channel. It includes the production yield for reactions not only at the incident beam energy, but also in the energy range below due to the energy loss of beam particles in the thick HOPG target. The cross section at the incident energy can then be obtained from the derivative $dY/dE$ of the thick-target yields measured in multiple small energy steps of $50$ keV in the center of mass \cite{notani2012}. This makes our effective target thickness $\sim 50$ keV in contrast to a few hundred keV of typical ``thin'' target experiments. The value of $dY/dE$ at a given energy was determined by fitting the yield at this energy together with the yields detected for the two neighboring energy steps using a second-order polynomial of $logY$ vs. $E$ \cite{notani2012}. Whereas this treatment is not possible at the edge of an energy range, a linear fit from one side is applied and results in larger uncertainties for $dY/dE$. The partial cross sections are derived from the extracted differential yield $dY/dE$ for each of the observed particle groups using the thin target equation,

\begin{equation} \label{eq_sigma_from_yield}
\sigma (E)=\frac{1}{\varepsilon}\frac{M_{T}}{f N_{A}}\frac{dE}{d\left ( \rho X \right )}\frac{dY}{dE}
\end{equation}

\noindent where $\varepsilon$ is the detection efficiency of measured $\gamma$ rays and charged particles in coincidence, $f$ is the molecular fraction of target nucleus, N$_{A}$ is the Avogadro constant, M$_{T}$ is the molecular weight of the target, and $dE/d(\rho X)$ is the stopping power calculated with SRIM \cite{SRIM}.

\begin{figure}[htb]
\centering
\includegraphics[width=0.5\textwidth]{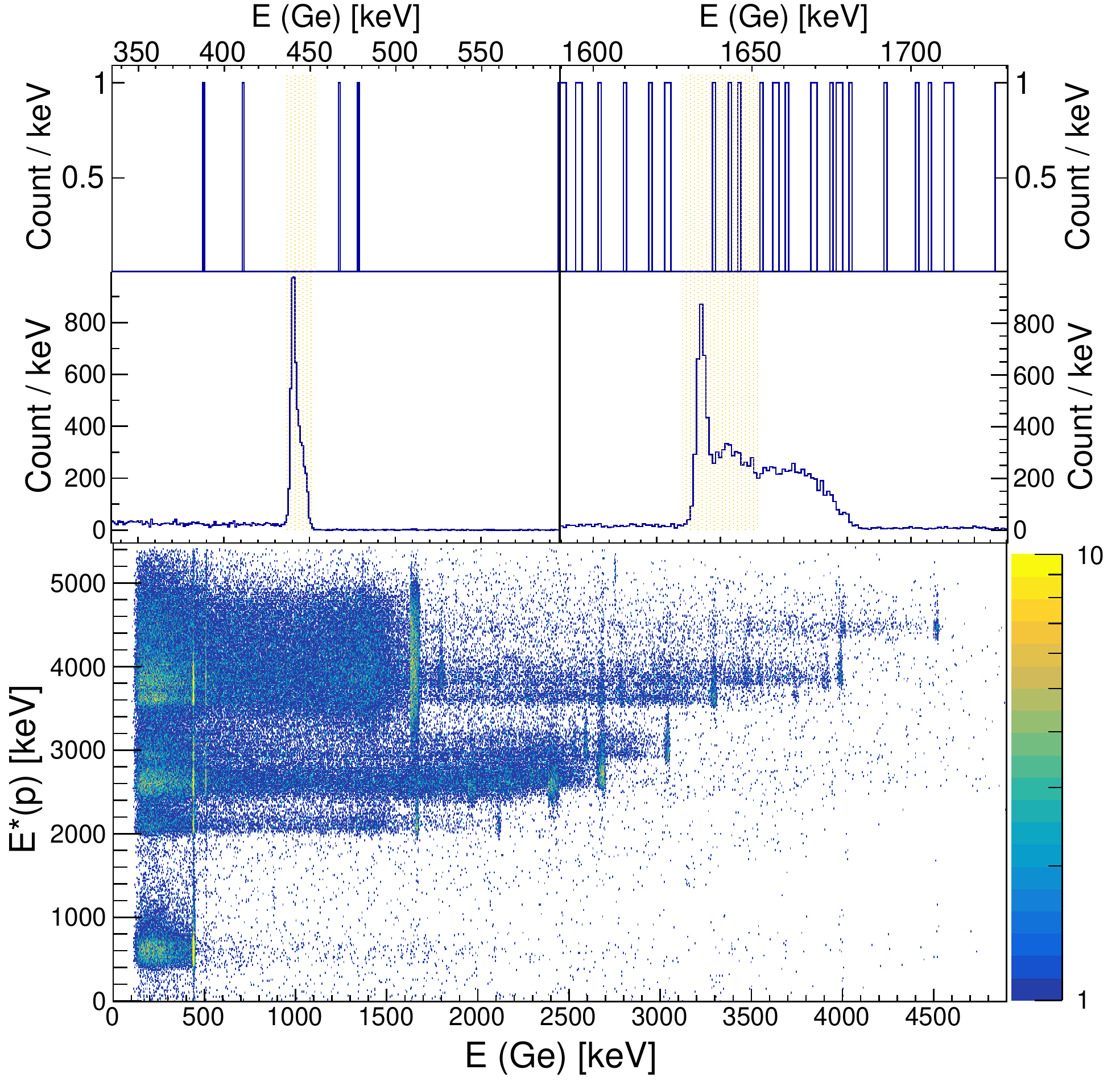}
\caption{
\label{fig_t2expge_run43}
A coincidence spectrum (bottom) for a typical run at E$_\text{beam}$=8.9 MeV between charged particles and gamma-rays is shown. The middle panels show the projections to the $\gamma$ energies and the Doppler effects for the $p_1$ (left) and $\alpha_1$ (right) channels, respectively, where the bump just outside the $\alpha_1$ shaded region is contributed from the $p_2$ channel. The top panels present the similar projections for the lowest energy at E$_\text{beam}$=4.4 MeV.
}
\end{figure}

Coincidence data between charged particles and $\gamma$-rays are shown in Fig.~\ref{fig_t2expge_run43} for a typical run at E$_\text{beam}$=8.9 MeV where $E(\text{Ge})$ is the energy of $\gamma$-rays detected in HPGe and $E^*(p)$ is the excitation energy of the residual nucleus calculated from the particle energy in SAND after kinematic corrections for the proton channel. The gated $p_1$ (left) and $\alpha_1$ (right) channel projections and corresponding Doppler effects are also shown in the middle panels for the same energy and in the top panels for E$_\text{cm}$=2.2 MeV of Fig.~\ref{fig_t2expge_run43}, respectively.

\begin{figure}[htb]
\centering
\includegraphics[width=0.5\textwidth]{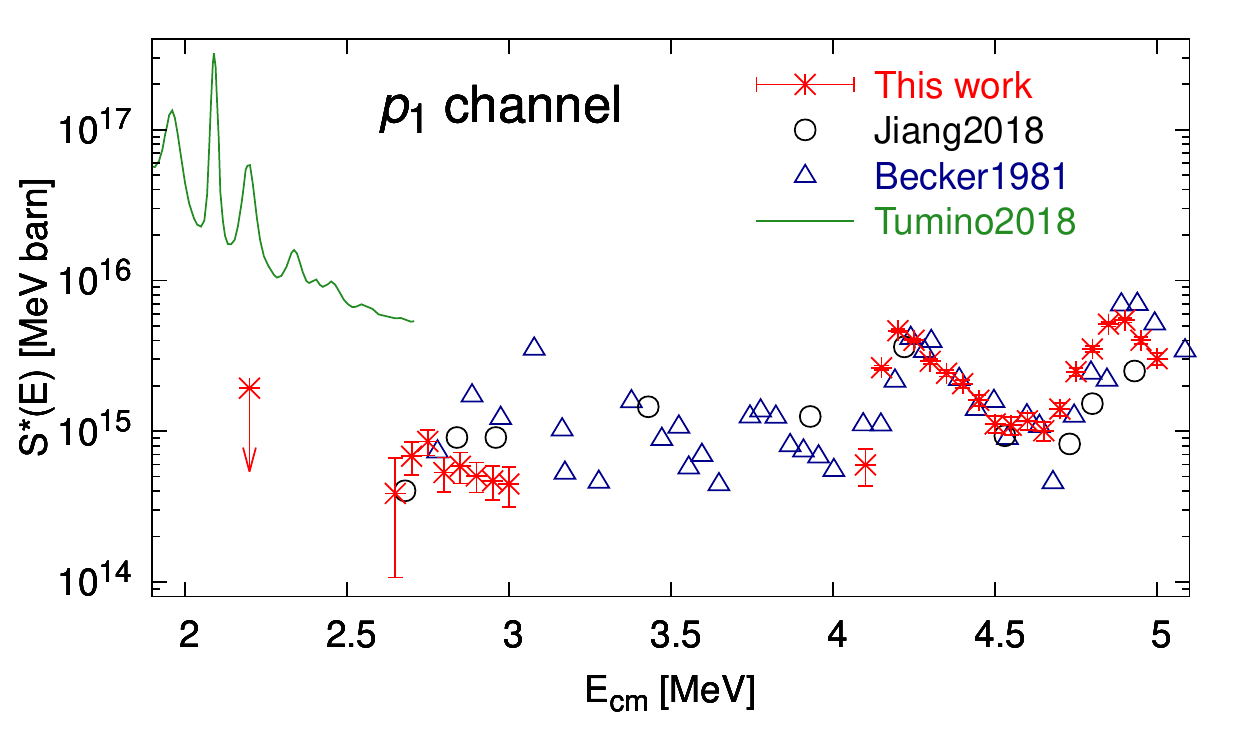}
\caption{
\label{fig_p1sfac}
The $p_1$ channel $S^*(E)$ factor for this work is shown and compared with previous work \cite{becker1981,tumino2018,jiang2018}.
}
\end{figure}

After applying the differential thick target method as presented in Eq. \ref{eq_sigma_from_yield}, we can obtain the $p_1$ cross section. For better comparison and presentation, it is customary to calculate the so-called $S(E)$ factor that removes removes the strong energy dependence caused by the Coulomb barrier. In particular, a special $S^*(E)$ factor for $^{12}$C+$^{12}$C commonly used in the literature \cite{barnes1985} is defined as follows,

\begin{equation} \label{eq_s_factor}
S^*(E)=\sigma(E)\cdot E\cdot  \exp(87.21/\sqrt{E}+0.46E)
\end{equation}
where $E$ is the center-of-mass energy in unit of MeV.
Fig. \ref{fig_p1sfac} shows the $p_1$ channel $S^*(E)$ factor for this work in comparison with previous work \cite{becker1981,tumino2018,jiang2018}. Other data that do not provide separate information on the $p_1$ channel are not shown. The error bars of our data shown in Fig. \ref{fig_p1sfac} are statistical. The additional systematic errors are about 5\% from the stopping powers and up to about 10\% from efficiency calibration and summing effects. We assume isotropic distributions in our measurement which can incur uncertainties of about 10\% for $p_1$ and 30\% for $\alpha_1$ based on previous measurements of angular distributions \cite{becker1981}. However, extreme cases and effects of possible correlations could significantly increase the uncertainties (e.g., about 20\% for $p_1$ and 60\% for $\alpha_1$ assuming an angular distribution of $\propto \cos^2(\theta)$).

The THM data (solid line) \cite{tumino2018} show about an order of magnitude higher values compared to our low energy data as can be seen in Fig. \ref{fig_p1sfac}. The trend towards lower energies seems to increase, which could be related to the mistreatment of Coulomb interactions as suggested by Ref. \cite{mukhamedzhanov2019}. In particular, the THM-predicted resonance at $E_\text{cm}=2.2$ MeV is in severe conflict with our upper limit derived from the absence of observed coincidence events as shown in Fig.~\ref{fig_t2expge_run43}. Towards higher energies ($>4$ MeV) and lower energies ($<2.8$ MeV) our data agree fairly well with the previous direct measurements including the resonant structures. At energies just below 3 MeV, the discrepancies with previous works \cite{becker1981,jiang2018} could be due to the systematic errors from angular distributions and / or the integrative effects of the effectively thick target yield affected by the resonance at 3.07 MeV.

\begin{figure}[htb]
\centering
\includegraphics[width=0.5\textwidth]{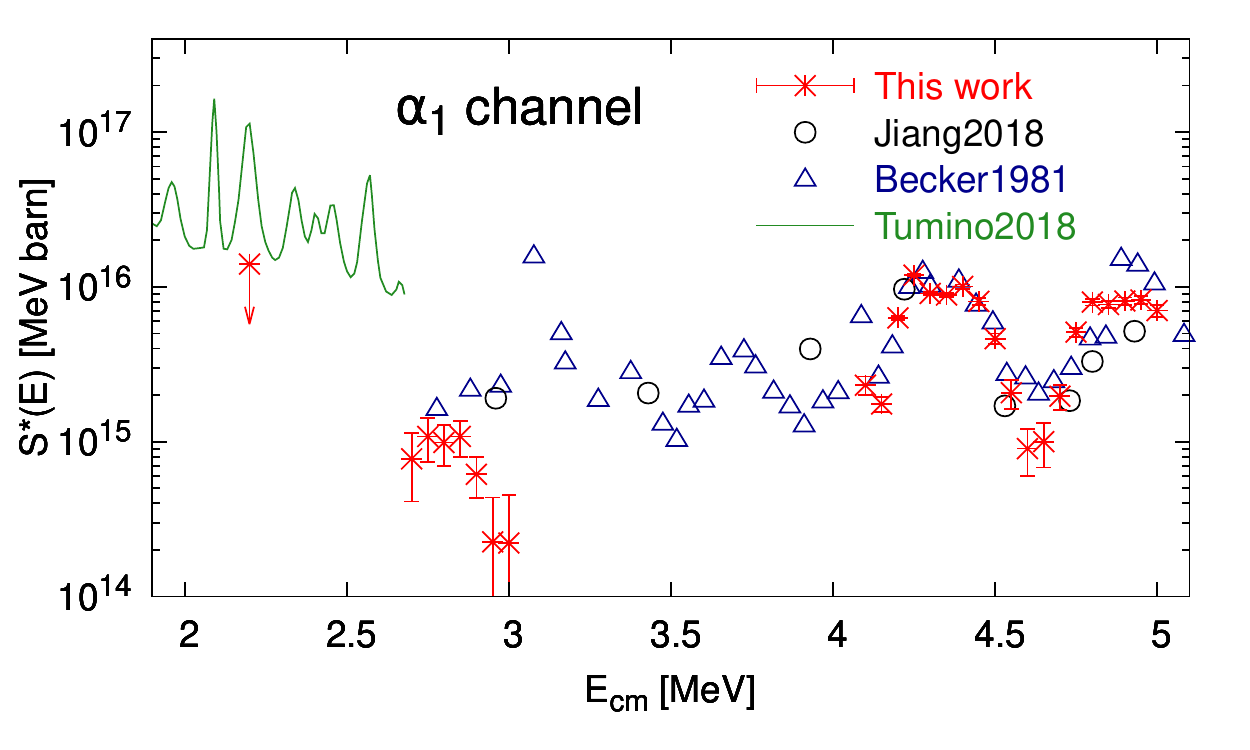}
\caption{
\label{fig_a1sfac}
The $\alpha_1$ channel $S^*(E)$ factor for this work is shown and compared with previous work \cite{becker1981,tumino2018,jiang2018}.
}
\end{figure}

Similarly, the $\alpha_1$ channel $S^*(E)$ factor for this work is presented in Fig. \ref{fig_a1sfac} in comparison with previous work \cite{becker1981,tumino2018,jiang2018}. The lowest data point at $E_\text{cm}=2.2$ MeV is estimated to be an upper limit assuming, rather conservatively, that all three observed coincidence events correspond to real events. Compared to the $p_1$ data, the $\alpha_1$ channel is more prone to background contamination even for coincidence measurements as shown in Fig.~\ref{fig_t2expge_run43}. As a matter of fact, the background level outside the coincidence gate is very similar to that within the gate suggesting that all the observed coincidence events may stem from background. If so, the estimate of the upper limit at the lowest energy can be much lower and may be close to the value of the $p_1$ channel. As such, our $p_1$ and $\alpha_1$ data sets have shown similar $S^*(E)$ values throughout the energy range.

The 1634 keV $\gamma$-ray peak from the $\alpha_1$ channel may potentially be contaminated by the 1636 keV $\gamma$-rays from either the $p_2$ channel or inelastic scattering of $^{23}$Na contamination within the target or anywhere the scattered beam can reach. This can cause severe background issues at low energies in measurements using the ``thin'' target approach. Previous data sets for the proton and alpha-particle channels have similar $S^*(E)$ factors at higher energies ($\gtrsim 3$ MeV) while differing significantly at lower energies ($\lesssim 3$ MeV). These enhanced alpha-particle channel values may be due to contributions of low energy $^{12}$C and $\alpha$ cluster resonances as suggested by the THM data or they could be simply due to the contamination as mentioned above. For example, the previously claimed resonance at $E_{cm}=2.14$ MeV \cite{spillane2007} (shown in Fig. \ref{fig_totsfac}) has a significantly larger contribution from the alpha-particle channels. The recent coincident measurement by the STELLA collaboration \cite{stella2019} shows similar $S^*(E)$ factors for both channels above $E_{cm}=3$ MeV while dramatically higher values in the $\alpha_1$ channel at energies below 3 MeV. In particular, the STELLA data provide a $p_1$ upper limit at $E_{cm}=2.2$ MeV similar to our measurement whereas they claim a much stronger $\alpha_1$ $S^*(E)$ factor at the same energy. We do not concur with that result. 

On the contrary, the $^{23}$Na contamination effect is largely canceled out in our differential thick target approach. In addition, the Doppler effect can shift the $p_2$ gamma rays by more than 30 keV as shown in the middle right panel of Fig.~\ref{fig_t2expge_run43}, which makes $\alpha_1$ and $p_2$ gamma rays better separated in our setup. Fig. \ref{fig_a1sfac} shows that our $\alpha_1$ data agree fairly well with existing data at high energies. Similar to the $p_1$ case, the differences at energies just below $E_{cm}=3$ MeV could stem from the uncertainties of angular distributions and / or the integrative effects of target thickness. Again, the data (solid line) from the indirect THM measurement are much higher than what our data show.

To compare our results with more of the available data on total $S^*(E)$ factor, we renormalized our data using the linear fit of the $p_1$-to-$p_{total}$ and $\alpha_1$-to-$\alpha_{total}$ ratios from the fairly complete data set of Becker \textit{et al.} \cite{becker1981}, which should be less sensitive to the above-discussed adverse effects. Unfortunately, such a normalization procedure translates to an uncertainty of up to a factor of two due to large fluctuations of the ratios in the data of Becker \textit{et al.} \cite{becker1981}. Nevertheless, our renormalized total $S^*(E)$ factor data (solid circles) without the additional normalization uncertainties are shown in Fig. \ref{fig_totsfac} in comparison with other available data. The above discussions for the $p_1$ and $\alpha_1$ channels can be applied here as well. The upper limit at the lowest energy can be lowered by more than a factor of two if we consider all the coincident counts observed in the $\alpha$ channel as background.
Our low energy data are lower than most of the previous works possibly due to significant contamination in the $\alpha_1$ channel of the previous measurements. Meanwhile, our data, consistent with the published thick-target yield of Zickefoose \textit{et al.} \cite{zickefoose2018}, agree with the singles measurement from Zickefoose's unpublished thesis work using the thick target approach \cite{zickefoose2011}, which was unfortunately hindered by much larger uncertainty and therefore not shown in Fig. \ref{fig_totsfac}.

\section{Conclusions}

New measurements of the $^{12}$C+$^{12}$C fusion reaction were conducted at Notre Dame using particle-$\gamma$ coincidence and differential thick target techniques which help to minimize possible contamination effects at low energies and reduce the uncertainty of target thickness integration due to the large energy loss associated with the large stopping power of low energy carbon beams. The new data provide a more reliable cross section and $S^*(E)$ factor compared to previous measurements. In particular, our results show strong disagreement with the recent THM data \cite{tumino2018} by more than one order of magnitude. This might be due to faulty normalization of the THM data to direct cross section measurements, but may also be due to a systematic error stemming from the mathematical treatment in converting the THM transfer reaction data into relative capture reaction cross sections. One possible way to reconcile the discrepancy is to re-analyze the THM data using a new normalization factor based on this work and taking into account the Coulomb effect as discussed in Ref. \cite{mukhamedzhanov2019}. 

\begin{acknowledgments}
We would like to thank Edward Stech and Daniel Robertson for their technical support on the operation of the 5U Pelletron accelerator at Notre Dame. We thank the STELLA collaboration for the permission to show their data in the total S-factor plot of Fig. \ref{fig_totsfac}.
This work is supported in part by the National Science Foundation under
grant No. PHY-1713857 and the Joint Institute for Nuclear Astrophysics (JINA-CEE, www.jinaweb.org), NSF-PFC under grant No. PHY-1430152. Also acknowledged is the support for the Mexico City Collaboration through a Grant from Notre Dame International (NDI) as well as partial support from CONACYT under Grant CB-01-254616.
\end{acknowledgments}


\bibliography{c12c12}

\end{document}